\documentclass{PoS}

\title{Coronal Sources of Impulsive Fe-Rich Solar Energetic Particle Events}

\ShortTitle{Coronal Sources of Impulsive Fe-Rich Solar Energetic Particle Events}

\author{\speaker{Stephen Kahler}\\
        {Air Force Research Laboratory} \\
        E-mail: \email{stephen.kahler@kirtland.af.mil}}

\author{Donald Reames\\
      \\{Institute for Physical Science and Technology, University of Maryland}} 
\author{Edward Cliver\\
    \\{National Solar Observatory and Air Force Research Laboratory}}

\abstract{We review recent work on 111 Fe-rich impulsive solar energetic ($\sim$ 3 MeV/nuc) particle (SEP) events observed from 1994 to 2013. Strong elemental abundance enhancements scale with A/Q, the ion mass-to-charge ratio, as (A/Q)$^{\alpha}$, where 2 $< \alpha <$ 8 for different events.  Most Fe-rich events are associated with both flares and coronal mass ejections (CMEs), and those with larger $\alpha$ are associated with smaller flares, slower and narrower CMEs, and lower SEP event fluences.  The narrow equilibrium temperature range required to fit the observed A/Q enhancements is 2.5--3.2 MK, far below the characteristic flare temperatures of $>$ 10 MK.  Only a small number of SEP events slightly outside this temperature range were found in an expanded search of impulsive Fe-rich events. Event characteristics are similar for events isolated in time and those occurring in clusters.  The current challenge is to determine the solar sources of the Fe-rich events.  Ambient coronal regions in the 2.5--3.2 MK range are broadly distributed both in and outside active regions.  We explore the possibility of acceleration from thermal plasmas at reconnecting current sheets in the context of observed standard and blowout jets.  Recent current sheet reconnection modelling provides a basis for the A/Q enhancements.}

\FullConference{The 34th International Cosmic Ray Conference,\\
		30 July- 6 August, 2015\\
		The Hague, The Netherlands}

\begin{document}

\section{Introduction}
A powerful tool in the study of SEP events is the composition of their elemental abundances, which has indicated the existence of two basic kinds of events [1, 2].  The large, gradual SEP events are composed of essentially an ambient coronal composition, but the smaller, more numerous impulsive SEP events were defined primarily by their distinctively non-coronal composition, with up to a 1000-fold increase of $^{3}$He/$^{4}$He and of heavy elements, such as (Z $\geq$ 50)/O. While the role of CME-driven shocks as progenitors of ambient SEP events has been well established, the connection of CMEs and flares to the impulsive SEP events and to their elemental abundance enhancements remains poorly defined.  Here we first review our recent study of the abundances of the elements He through Pb in Fe-rich impulsive solar energetic-particle (SEP) events with measurable abundances of ions with atomic number Z $>$ 2 observed on the \textit{Wind} spacecraft.  We related the Fe-rich impulsive SEP events to solar events [3].  In [4] we compared the event-to-event variations of source plasma temperatures and power-law abundance enhancements with properties of associated CMEs and flares.  Recent new work on these Fe-rich events is briefly described and we conclude with some speculation about impulsive SEP production in jets.  There are two basic goals of the work: to establish the coronal regions and conditions of the SEP source plasma population and to determine the mechanism and observable properties of the acceleration process(es).

\section{SEP Element Abundance Observations and Analysis}
The Fe-rich event samples were taken from a matrix of Ne/O versus Fe/O ratios at 3 MeV/nuc over 8-hr intervals observed from 1994 to 2013 on the $Wind$ spacecraft.  The time periods were examined with 15-min averages for discrete events with well defined onsets and durations.  A total of 111 Fe-rich events were found and their CME, flare, and decametric-hectometric (DH) type III radio burst associations tabulated [3].  Candidate DH type III bursts found for 95 of the 111 events provided good event onset fiducials.  We calculated all elemental abundance enhancements of He through Pb at 3--5 MeV/nuc after establishing by comparing the event Fe and O spectral indices that there was essentially no energy dependence of the abundance enhancements over the range 2--15 MeV/nuc.  

Abundance enhancements are better organized by elemental A/Q, mass to charge ratios, than by atomic number Z. Q is temperature dependent, and for this we assumed a coronal range of 2.5--3.0 MK [5] shown as the pink band of Figure 1.  The mean enhancements normalized to O are shown in the log-log plot of Figure 2, in which a power-law fit with exponent $\alpha$ = 3.6 represents the data well.

\begin{figure}
%fig 1
\centering
\noindent\includegraphics[width=15pc]{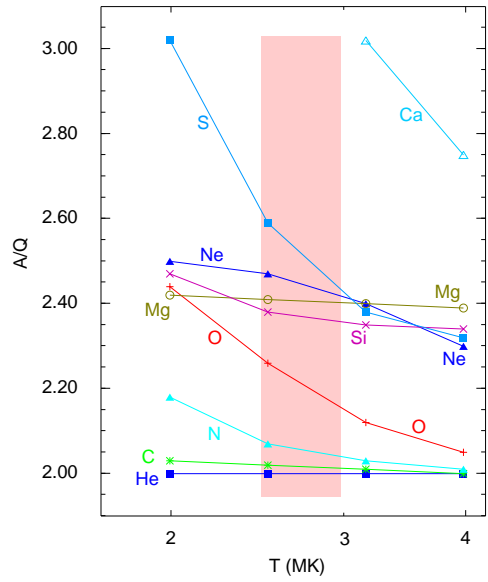}
\caption{A/Q versus equilibrium temperature in MK for low-Z elements.  Only in the pink band of likely temperatures do Si, Mg, and Ne show consecutively increasing A/Q. Figure 7 of [3].}

\end{figure}

%\twocolumn
\begin{figure}
%fig 2
\centering
\noindent\includegraphics[width=15pc]{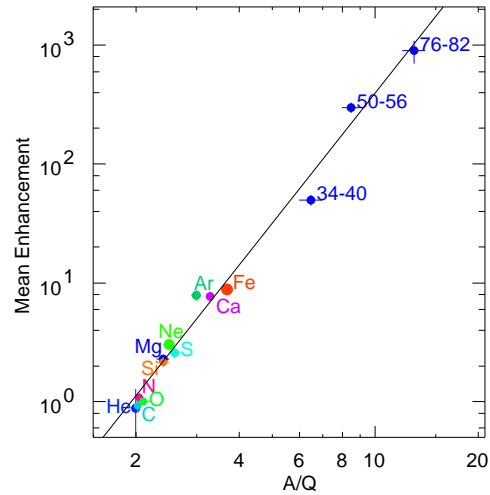}
\caption{Mean enhancements of elemental abundances versus elemental A/Q relative to coronal abundances for the 111 Fe-rich SEP events.  Upper data points are elements grouped by atomic numbers Z.  The slope of the best-fit line $\alpha$ is 3.64 $\pm$ 0.15. Figure 8 of [3].}

\end{figure}

\begin{figure}
%fig 3
\centering
\noindent\includegraphics[width=15pc]{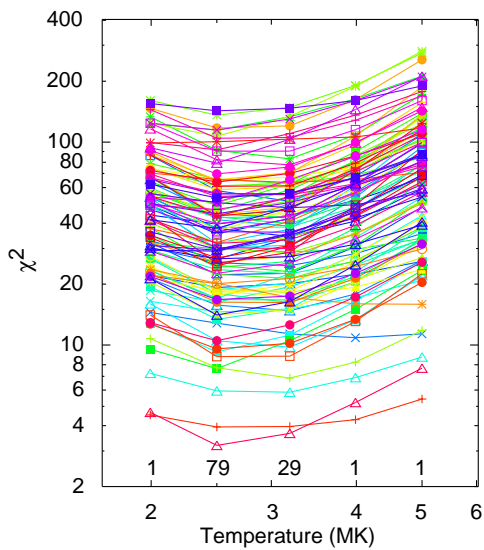}
\caption{$\chi^{2}$ vs. T is shown for all 111 impulsive SEP events using different colors and symbols for each event.  The number of events with $\chi^{2}$ minima at each temperature is shown along the bottom of the panel. From [4].}

\end{figure}

\begin{figure}
%fig 4
\centering
\noindent\includegraphics[width=15pc]{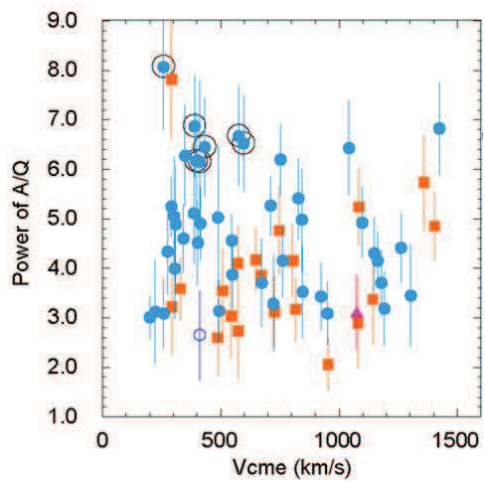}
\caption{The power $\alpha$ versus the associated CME speed. Temperature color coding: 2.5 MK (blue circles); 3.2 MK (red squares). Circled events are those identified as He-poor in RCK1.  From [4].}

\end{figure}

\begin{figure}
%fig 5
\centering
\noindent\includegraphics[width=15pc]{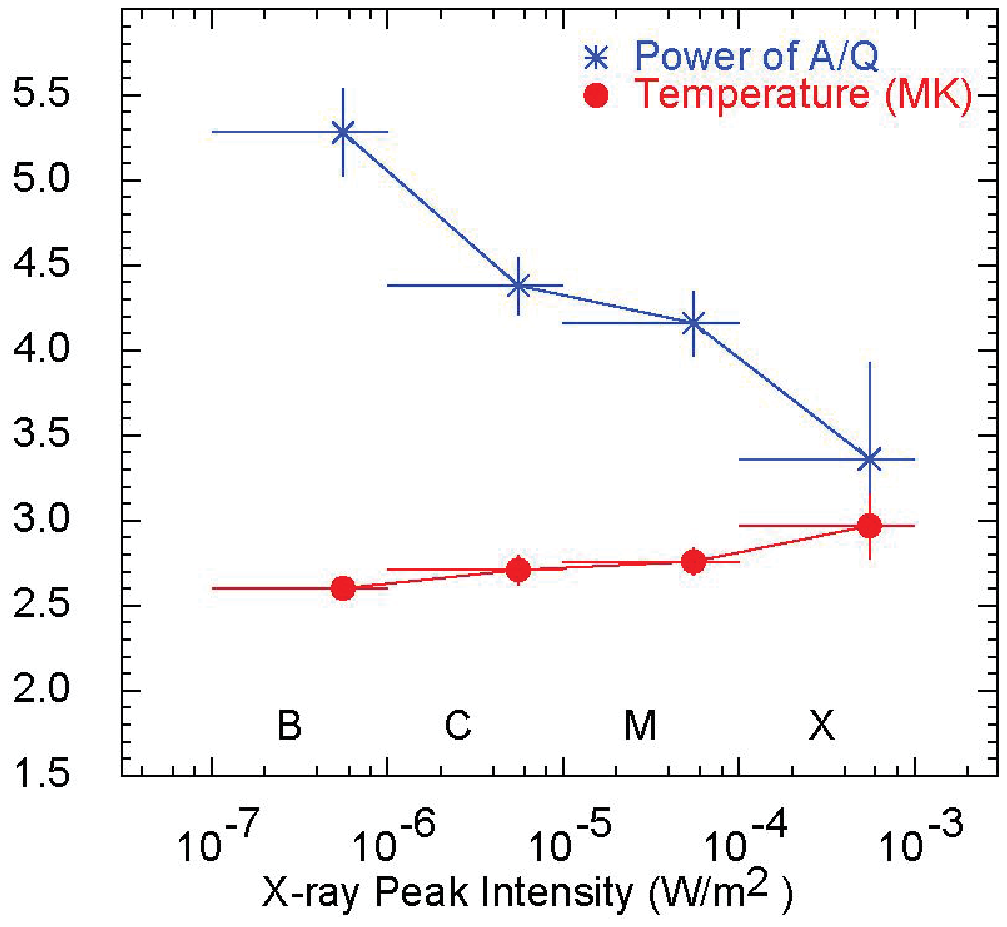}

\caption{Variation of the power $\alpha$ and of coronal temperature (in MK) versus X-ray flare peak size. From [4].}

\end{figure}
%\onecolumn

We [3] found 66 associated CMEs for the 96 SEP events (69\%) with available LASCO data.  Nearly all (61/66) were from the western hemisphere, as were the associated solar flares [3].  There was a significant correlation (r = 0.41) between the event O fluences and CME speeds.  In general the larger fluence SEP events were associated with wider and faster CMEs, and the median CME speed of 597 km s$^{-1}$ and width of 75$^{\circ}$~exceed those of all CMEs, but are much less than those of CMEs associated with gradual SEP events.  The 30 SEP events for which we found no associated CMEs are excluded from these median values.  The trend for smaller fluence SEP events to be associated with narrower and slower CMEs is consistent with an association of a CME for every SEP event, below a LASCO detection threshold for small events but perhaps driving a SEP-producing shock in the largest fluence SEP events.

How do the SEP elemental abundance enhancements correlate with CMEs and flares?  The event enhancements are a product of both the source temperature T, which determines the A/Q of each element (Figure 1), and the acceleration mechanism, which appears well described as a power law of A/Q (Figure 2).  To extract the optimum values of T and $\alpha$, we [4] did $\chi^{2}$ fits for five assumed T values over the range 1--5 MK for each of the 111 events and selected the minimal $\chi^{2}$ fit to determine T (Figure 3).  This temperature range is required to obtain increasing A/Q and the higher observed abundance enhancements for decreasing Z among elements Ne, Mg, and Si (Figure 1).  For nearly all events T = 2.5 or 3.2 MK.  With the optimal T fit, we also derived the associated exponent $\alpha$ of the (A/Q)$^{\alpha}$ fit, and that gave us our elemental abundance enhancement parameter for each event.  In Figure 4 we compare $\alpha$, color-coded for T, of each event with the associated CME speed.  The correlations with CME speeds and widths (not shown) are both weak.  Many events with the steepest fits ($\alpha >$ 6) are He-poor, i.e., He/O is low because A/Q of O is well above 2 in the low-T range (Figure 1).

We show the separate trends of $\alpha$ and T with increasing X-ray flare size in Figure 5.  The predominate trend is for decreasing $\alpha$ with increasing flare size.  We also found trends of decreasing $\alpha$ with harder O energy spectra and with larger logs of O event fluences.  To summarize our [4] statistical comparisons, we find first that both T and $\alpha$ vary from event to event.  The event abundance enhancements of the elements He through Pb, as measured by $\alpha$, decrease with: increasing event O fluences, coronal temperatures, X-ray flare sizes, CME speeds and widths, and flatter O energy spectra.

\section{Recent Analysis}
\subsection{Search for Further Properties of Fe-Rich SEP Events}
The 111 impulsive events of [3] and [4] were selected to be Fe-rich, specifically (Fe/O)/0.131 $>$ 4. Does that selection criterion preclude events outside the range of 2.5--3.2 MK shown in Figure 3?  To look for cooler impulsive events, we selected He-poor events, those with low He/O abundance enhancements due to the increased A/Q of O (Figure 1).  That search yielded 11 new events, but our $\chi^{2}$ analysis of T (Figure 3) yielded only 3 new events with T = 2 MK, with the remaining values in the predominate range of 2.5 or 3.2 MK.  Figure 1 also instructs us how to look for possible hotter impulsive events.  The enhancement of Ne no longer exceeds that of Si, and those of S and Si should be comparable.  A search on that basis yielded only 4 new events, of which 3 had T = 4 MK as best fits.  Thus for the expanded list of 126 impulsive events the temperature range of 2--4 MK still dominates.  The $\alpha$ values range from 2 to 8 with a mean of 4.5 and standard deviation of 0.74.

\subsection{Fe-Rich SEP Events in Time}

The impulsive event list (RCK1) shows that many events are clustered in time.  If we consider for each event the time interval from its closer neighbor in the list, we find that 44 of the 111 events are separated from each other by $\le$ 1.5 days.  Do the clustered events show any different behavior
which might reflect some systematic depletion of hi-Z elements or a change in acceleration mechanism?  We considered event O fluence, $\alpha$, and T as functions of the event separation times and found no correlations.  This implies a lack of any kind of observed physical coupling between events closely paired in time. We also looked for any solar cycle dependence of the Fe-rich events.  The event numbers trend with solar activity, with a possible tendency for lower $\alpha$ and higher T at solar maximum.

\section{Discussion}

Although the impulsive SEP events are strongly associated with flares, the consistent 2--4 MK (log 6.3--6.6) ionization temperatures deduced for the SEP ions precludes their acceleration from flare-heated plasma, which usually exceeds 10 MK [6,7].  Further, the elemental composition of flares has been found to be characteristic of the deep chromosphere [8] and therefore inconsistent with a coronal plasma [9] at any temperature.   Evidently the ions of 2 to 20 MeV/amu that we observe in space are accelerated from active-region (AR) plasma on open magnetic field lines near the flare, but not from the flare plasma itself.  EUV AR observations show a hotter core, composed of closed field lines, surrounded by a cooler diffuse region [10], perhaps of mostly open field lines and likely sources of the Fe-rich SEP events. Emission-measure distributions of the core/diffuse boundaries of two ARs showed broad peaks in the log 6.3--6.6 temperature range [10], but that range is also well represented in emission measure distributions deduced for AR cores [11, 12].  On the Atmospheric Imaging Assembly (AIA) of the Solar Dynamics Observatory the 2--4 MK range is best imaged with the 335 \AA~Fe XVI channel [13], which shows both AR and diffuse coronal emission (readily viewed at http://sdo.gsfc.nasa.gov/data/).  All AR complexes appear to be candidate source regions for production of the Fe-rich impulsive SEP events based on their appropriate emission measure profiles.

To account for the flare and CME associations the impulsive SEPs must be accelerated near an AR and ejected along open field lines.  Rust et al.[14] used \textit{Yohkoh} soft X-ray images to study the solar sources of 25 impulsive injections of near-relativistic electrons detected at 1 AU.  They found small open regions (AR coronal holes) adjacent to flaring ARs in 23 cases (4 of which are Fe-rich events of [3]) and compared their AR magnetic geometry to the reconnection model of [15].  Their 2-D model of interchange reconnection between a flaring loop and adjacent open fields allowed an open field path for accelerated particles to escape the corona.  

Recent detailed observations of coronal jets, much smaller events best observed from coronal holes, can provide some useful guidance here.  Moore et al.[16] identified two kinds of X-ray jets in polar coronal holes, standard and blowout jets, which they interpreted in terms of the basic Shibata (e.g., [17]) model of interchange reconnection between an open magnetic field and a magnetic loop or arcade.  The increasing pressures between the open field and emerging magnetic arcade provide the magnetic free energy to drive the initial interchange reconnection that results in a standard X-ray emitting jet.  If the arcade field has sufficient twist and shear, that resulting magnetic energy can drive a subsequent eruption of the arcade, which reconnects both internally and with open field to become the blowout jet [18], visible as a later cooler EUV component [19].  The cool component shows a helical, or axial, motion [19, 20], interpreted in terms of an unwinding of the twist of the erupting arcade.  If we associate our Fe-rich SEP events with jets, then the high association of the events with CMEs argues that we are dealing with large-scale blowout jets, which are also characterized by lateral expansions in width by a factor of 2 or 3 [20].

The acceleration of the impulsive SEPs must occur early, before significant plasma heating can take place to change the A/Q values of the ions.  In the jet scenario, acceleration must occur in the initial stage of reconnection that leads to the standard jet, although the jet plasma itself may be heated.  Recent models of standard and blowout jets [21, 22, 23, 24] have shown temperatures of the standard jet component propagating on open field lines to be in the 2--4 MK or higher ranges.  The implication of this picture is that the blowout jet, perhaps manifested as a CME, is not directly connected to the earlier particle acceleration phase leading to the standard jet, but rather is a later phase of an eruptive event.

Recent modelling work on ion acceleration at reconnecting current sheets has shown how an A/Q bias can arise when reconnection occurs within a strong guide field [25, 26].  The ions must behave as pick-up ions in the reconnection exhaust, and the threshold for that condition is A/Q $>$ k$\beta^{0.5}$, where k is the inverse reconnection rate and $\beta$ is the magnetic to plasma pressure ratio just upstream of the exhaust.  $\beta$ gradually declines as the reconnection proceeds [26], and that allows ions of progressively lower A/Q to be accelerated.  Calculated energy gains were very modest ($\sim$ 25 keV/nuc), but the resulting energized population could be further accelerated in first-order Fermi interactions with multiple magnetic islands predicted to occur in flares [25].  

We have broadly sketched a scenario in which coronal thermal ions are accelerated with an A/Q enhancement in a localized magnetic reconnection region with access to open field lines and accompanied by a subsequent magnetic arcade ejection in the form of a blowout jet or CME.  Although we can not rule out some contribution from shock acceleration in the Fe-rich events, reconnecting current sheets seem to provide the most promising explanation for the events of the study.

\textbf{Acknowledgements}
S. Kahler was funded by AFOSR Task 15RVCOR167.  CME data were taken from the CDAW LASCO catalog.  This CME catalog is generated and maintained at the CDAW Data Center by NASA and The Catholic University of America in cooperation with the Naval Research Laboratory. SOHO is a project of international cooperation between ESA and NASA.

\end{document}